# Binary Change Guided Hyperspectral Multiclass Change Detection

Meiqi Hu, *Graduate Student Member*, *IEEE*, Chen Wu, *Member*, *IEEE*, Bo Du, *Senior Member*, *IEEE*, and Liangpei Zhang, *Fellow*, *IEEE*

*Abstract*—Characterized by tremendous spectral information, hyperspectral image is able to detect subtle changes and discriminate various change classes for change detection. The recent research works dominated by hyperspectral binary change detection, however, cannot provide fine change classes information. And most methods incorporating spectral unmixing for hyperspectral multiclass change detection (HMCD), yet suffer from the neglection of temporal correlation and error accumulation. In this study, we proposed an unsupervised Binary Change Guided hyperspectral multiclass change detection Network (BCG-Net) for HMCD, which aims at boosting the multiclass change detection result and unmixing result with the mature binary change detection approaches. In BCG-Net, a novel partial-siamese united-unmixing module is designed for multi-temporal spectral unmixing, and a groundbreaking temporal correlation constraint directed by the pseudo-labels of binary change detection result is developed to guide the unmixing process from the perspective of change detection, encouraging the abundance of the unchanged pixels more coherent and that of the changed pixels more accurate. Moreover, an innovative binary change detection rule is put forward to deal with the problem that traditional rule is susceptible to numerical values. The iterative optimization of the spectral unmixing process and the change detection process is proposed to eliminate the accumulated errors and bias from unmixing result to change detection result. The experimental results demonstrate that our proposed BCG-Net could achieve comparative or even outstanding performance of multiclass change detection among the state-of-the-art approaches and gain better spectral unmixing results at the same time.

## I. Introduction

KNOWN for the high spectral resolution, hyperspectral image (HSI) provides tremendous spectral information for object discrimination and has made advanced development in hyperspectral classification [1, 2], target detection [3, 4], anomaly detection [5, 6], denoising [7, 8], spectral unmixing [9, 10] and change detection [11, 12]. Among various researches of hyperspectral images, change detection has been one of the hottest remote sensing application topics in the past decades, aiming to detect the change information of two remote sensing images acquired in the same geographical location on different times [13, 14], and has been widely applied in land cover and land usage change detection [15, 16], disaster emergency and assessment [17, 18], dynamic urban monitoring and management [19, 20], etc.

In general, hyperspectral change detection is categorized as hyperspectral binary change detection (HBCD) [21] and hyperspectral multiclass change detection (HMCD) [22]. HMCD has long been a challenging research work. Compared with HBCD, which requires to detect the changed area, HMCD not only needs to detect the change area and the unchanged area, but also demands to identify the number of changes and distinguish different kinds of change transformation associated with variant land-cover transitions [23, 24]. By contrast, HMCD offers more detailed change information.

The most straightforward solution to HMCD, named post classification comparison (PCC) [25], is to classify the HSIs separately and then compare the multi-temporal classification map to get the multiclass change detection result. However, abundant training samples are desired for accurate classification. Moreover, either classification error of the multiple HSI classification maps can cause plenty of erroneous change indications and false positive changes. Then Du et al.[26] proposed an unsupervised or semi-supervised HMCD method incorporating the spectral unmixing to detect multiclass changes, called post unmixing comparison (PUC) [27], where the independent unmixing is conducted on each image and corresponding abundance vectors of each pixel are compared to acquire binary and multiclass changes results. Generally, spectral unmixing is the process of decomposing the spectral signature of a mixed pixel into a set of pure materials (called endmember) and their corresponding proportions (called abundance) [28, 29], and is a powerful tool to the mixed pixel problem due to coarse spatial resolution of HSI. Much of the interest in the utilization of unmixing for HMCD is associated with the support of sub-pixel information provided by the abundance. Liu et al. proposed a novel multitemporal spectral unmixing (MSU) [30] approach which extracted the multi-temporal endmembers (MT-EMs) from the stacked hyperspectral images, and introduced a change analysis strategy to distinguish change MT-EMs from no-change MT-EMs. Song et al. developed a recurrent three-dimensional fully

This work was supported in part by the National Natural Science Foundation of China under Grant T2122014 and 61971317, and in part by the Natural Science Foundation of Hubei Province under Grant 2020CFB594, and in part by the Fundamental Research Funds for the Central Universities. *(Corresponding author: Chen Wu.)*

M. Hu, C. Wu and L. Zhang are with the State Key Laboratory of Information Engineering in Surveying, Mapping and Remote Sensing, Wuhan University, Wuhan 430079, China (e-mail: meiqi.hu@whu.edu.cn; chen.wu@whu.edu.cn; zlp62@whu.edu.cn).
B. Du is with the School of Computer Science, Wuhan University, Wuhan 430072, China (email: gunspace@163.com).

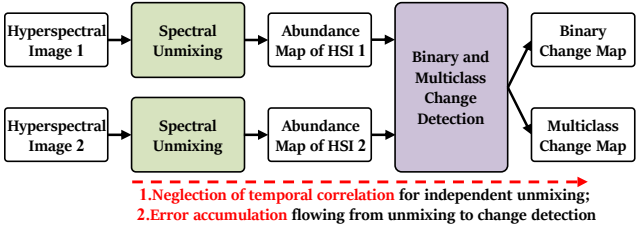

**Fig. 1.** Traditional framework of incorporating spectral unmixing for hyperspectral multiclass change detection.

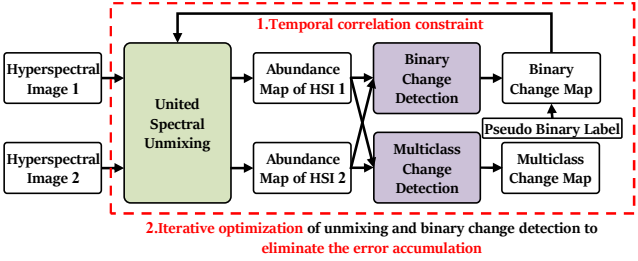

**Fig. 2.** Proposed binary change guided multiclass change detection framework.

convolutional network (Re3FCN) for HMCD [31], where a multiclass sample generation method combined with endmember extraction was proposed for network training. Hamid et al. put forward a new rule of detecting binary and multiclass changes by identifying the pure changed pixels and mixed changed pixels hierarchically from the abundance vectors pairs [32]. Guo et al. [33] represented an original temporal spectral unmixing model by using the current temporal HSI and the endmembers of the previous to obtain the endmember of the current temporal HSI, then the binary and multiclass changes were detected from the abundance maps. By assembling the endmember or abundance maps of unmixing results, the approaches mentioned above provide a number of excellent multiclass change detection solutions. Nonetheless, most of the methods suffer from error accumulation when the binary and multiclass change detection results are directly computed from the abundance maps (presented in **Fig. 1**). The accuracies of the binary and multiclass change detection results are significantly dependent on the unmixing results. In such case, the error accumulation also leads to an undermined multiclass change detection result.

It is found that the binary change detection result computed by spectral unmixing result reflects the performance of unmixing to some extent and contains the error accumulation and bias between unmixing and change detection. Can we start with improving the binary change detection result to further accomplish collaborative optimization between unmixing and change detection? In this paper, we propose a binary change guided multiclass change detection approach to solve the problem of error accumulation, as shown in **Fig. 2**. We design a temporal correlation constraint directed by the pseudo-labels of binary change detection result to guide the unmixing process from the perspective of change detection. Actually, the binary change result of multi-temporal images is a token of temporal correlation. That is, if the pixel at the same location at two phases has not changed, the abundance of the two pixels are likely to be similar; if the pixel has changed, the abundance of them are likely to be quite different. Specifically, given the reliable pseudo-labels from binary change detection result, the gaps between the binary change detection result calculated from the spectral unmixing results and that of pseudo-labels indicate the accumulated error and bias passing from the unmixing process to the change detection. Then the gaps are minimized through back propagation, which act over the unmixing process to boost the abundance of the unchanged pixels more similar and that of the changed pixels more accurate. Driven by the iterative optimization of unmixing results and binary change detection results, the error accumulation from the unmixing process to the change detection is eliminated, and high-accuracy binary and multiclass change detection result are accomplished. Meanwhile, more accurate multi-temporal spectral unmixing results can be obtained.

Inspired by this idea, an unsupervised multi-task learning neural network, Binary Change Guided hyperspectral multiclass change detection Network (BCG-Net), is put forward for HMCD. Firstly, we developed a novel partial-siamese united-unmixing module for multi-temporal spectral unmixing, where the scaled spectral attention block is designed to dig diverse spectral features, especially for the emphasis on the distinguished spectral characteristics of different ground objects. The sum-to-one constraint and non-negative constraint are integrated into the neural network perfectly, making it convenient to conduct unmixing in an unsupervised way. The bi-temporal HSIs are both fed into the network to acquire the abundance maps simultaneously. Secondly, the multi-temporal abundance vectors are input into the designed temporal correlation module to acquire the binary change information, testifying the performance of unmixing from the point of view of change detection. Then the unmixing and change detection are iteratively optimized to eliminate the bias and the accumulated error, yielding greater multiclass change detection result.

The main contributions of this article are summarized as follows:

1) To solve the accumulated errors problem, we propose an innovative thought of using binary change detection result to guide multiclass change detection, where the temporal correlation constraint is designed to instruct the unmixing process, boosting the abundance of the unchanged pixels more similar and that of the changed pixels more accurate. And iteratively optimization of the unmixing and change detection are designed to clear the impact of error accumulation and bias, contributing to greater binary and multiclass change detection results.

2) Based on the proposed thought, we put forward an outstanding unsupervised multi-task learning network, named Binary Change Guided hyperspectral multiclass change detection Network (BCG-Net). This network develops a novel united-unmixing module for multi-temporal spectral unmixing and a groundbreaking temporal correlation module for detecting the binary change detection based on the abundance vectors, which also serves as the temporal correlation constraint operated on the united unmixing process.

3) We conducted abundant experiments on the proposed method. The experimental results showed that our method could gain comparable or even outstanding performance among the state-of-the-art methods, demonstrating the effectiveness of proposed BCG-Net. And the discussion about the effect of temporal correlation constraint on the united unmixing and the change detection indicated that the temporal correlation constraint really worked on both the temporal unmixing result and the binary and multiclass change detection result.

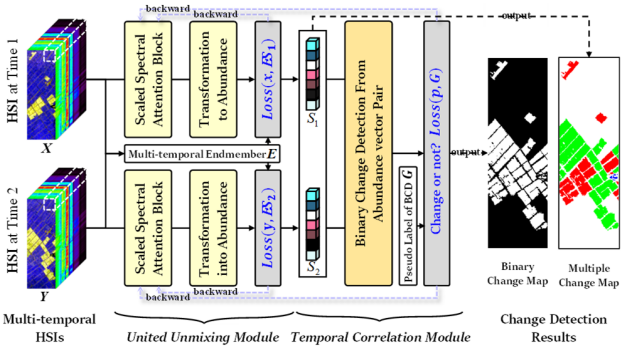

**Fig. 3.** Architecture of the proposed BCG-Net. The network consists of united unmixing module and temporal correlation module. And the temporal correlation module verifies the performance of the unmixing result in the view of change detection, placing a temporal correlation constraint on the united unmixing module.

The rest of the article is organized as follows. The detailed description of the proposed BCG-Net will be provided in Section II. And the data description and experiment results will be represented in Section III. Section IV will discuss the effect of temporal correlation constraint on spectral unmixing and change detection result. The conclusion of this paper will be given In Section V.

## II. METHODOLOGY

### A. General Framework

The general framework of BCG-Net is represented in **Fig. 3**. Mathematically, denote the bi-temporal HSIs at time 1 and time 2 by $X \in \mathbb{R}^{H \times W \times C}$ and $Y \in \mathbb{R}^{H \times W \times C}$, where $H$, $W$, as well as $C$ represent the height, width and the channel of the HSI, respectively; let $S_1$ and $S_2$ represent the abundance vectors of the center pixel of corresponding HSI, and the final output of the BCG-Net $p$ refers to the change probability of input center pixel. In addition, the multi-temporal endmember matrix $E$ extracted from the multi-temporal HSIs covers all the endmembers of the test HSIs. The pseudo-labels of binary change detection $G$ are obtained from the pre-detection binary change detection result, which are supposed to be reliable since the last decades have seen extensive and promoted progress of binary change detection methodology. As shown in **Fig. 3**, the BCG-Net consists of two parts, namely, the united unmixing module (UU-Module) and temporal correlation module (TC-Module). A pair of patch blocks centered on a certain pixel of the bi-temporal images are first fed into the UU-Module, which is tailored to get the corresponding abundance vectors. The outputs of the UU-Module are then fed into the TC-Module, which is designed to acquire the binary change information of the pixel. And the result is then compared with the given binary pseudo-label. The united unmixing result is qualified if the binary change result is consistent with the given binary label. Otherwise the unmixing result would be optimized further and the binary change result would be further compared with the given label. During the back propagation, the TC-Module takes a role of temporal correlation constraint on the UU-Module in the view of change detection, where the unmixing process is optimized further upon boosting the coherence of the abundance vectors of those unchanged objects to decrease the false change alarms, and more accurate abundance vectors for those changed objects. Consequently, better binary change result is outputted from the network and the multiclass change result is produced by the optimized united unmixing result.

### B. United Unmixing Module

The UU-Module aims at extracting the corresponding abundance from bi-temporal HSIs by united spectral unmixing. The UU-Module composes of two blocks, where Scaled Spectral Attention Block is designed to extract the spatial and multi-scaled spectral features from the input and the Transformation into Abundance Block to convert the extracted features into abundance. The UU-Module is in a partial-siamese structure, taking the similarity of bi-temporal inputs into consideration to generate accurate multi-temporal abundance.

For a single hyperspectral image $X$, the details of the single-branch UU-Module are shown **Fig. 4**. The hyperspectral patch $X_{patch} \in \mathbb{R}^{m \times m \times C}$ centered on a pixel $x \in \mathbb{R}^{C \times 1}$ is first processed by the 3×3×1 three-dimensional convolutional layer (3Dconv) [34] and 3×3×3 3Dconv separately, followed by two 3×3×3 3Dconvs respectively. The 3Dconv is able to extract spatial and spectral features simultaneously. The spectral kernel with size only equal to one is introduced to extract the band-by-band spectral features, while the spectral kernel with size equivalent to three is employed to dig the spectral feature of contiguous spectral channels. The different spectral kernel sizes are designed to extract spectral features of different scales. Furthermore, the efficient channel attention (ECA) [35] block (shown in **Fig. 4** (b)) is introduced to obtain the spectral attention after the two 3Dconvs, aiming at capturing the cross-channel interaction and stressing the discriminated spectral features. The one-dimensional convolutional layer designed in the ECA learns channel attention faster and more efficiently than fully connection layer of the Squeeze-and-Excitation Network [36] does. And the local kernel size is adaptively determined by a function of channel dimension. The two feature extraction branches focus on different scaled spectral features, which are beneficial to distinguish the landmark spectral features of different ground objects in spectral unmixing. Another 3×3×1 3Dconv is adopted after the extracted diverse features are fused by the pointwise concatenation of two branches. Additionally, no max pooling layer is used in the Scaled Spectral Attention Block. The Transformation into Abundance Block converts the spatial and spectral features into abundance vector of the center pixel using 1×1 two-dimensional convolutional layer (2Dconv). It is noted that ReLU [37] activation function is adopted to satisfy the non-negative constraint of abundance as the last layer. The length of the output abundance vector is equivalent to the number of the endmembers of multi-temporal HSIs. And the sum-to-one function $F$ is conducted after the ReLU activation, designed to meet the sum to one constraint of abundance vector. Given a vector $S = [s_1, s_2, ..., s_K]^T \in \mathbb{R}^{K \times 1}$, $F(s_i)$ is defined as:

$$F(s_i) = \frac{s_i}{\sum_{i=1}^{K} s_i + \varepsilon}, i = 1, 2, ..., K \quad (1)$$

where $\varepsilon$ is a very small value to avoid the computation disorder. The output abundance vector $S_1 \in \mathbb{R}^{K \times 1}$ of the first single-branch UU-Module can be expressed as:

$$S_1 = U_1(X_{patch}) = F(C_5(C_4(C_3(Feature_X))))$$
$$Feature_X = ECA_1(C_{12}(C_{11}(X_{patch}))) + ECA_2(C_{22}(C_{21}(X_{patch}))) \quad (2)$$

where $K$ depicts the total number of all multi-temporal endmembers; the $U_1$ refers to the single branch of UU-Module with input as hyperspectral $X$; $C_{11}$, $C_{12}$, $C_{21}$, $C_{22}$ and $C_3$ are all 3Dconvs of Scaled Spectral Attention Block; $ECA_1$ and $ECA_2$ represent the ECA block, while $C_4$ and $C_5$ are the 2Dconv of Transformation into Abundance Block. Then the predictive spectrum of the center pixel $\hat{x} \in \mathbb{R}^{C \times 1}$ can be obtained according to the linear mixed model [38], by the linear combination of the multi-temporal endmember matrix $E \in \mathbb{R}^{C \times K}$ and the output abundance vector $S_1$, and the formulation is expressed as follows:

$$\hat{x} = E \cdot S_1 \quad (3)$$

where the multi-temporal endmember matrix $E$ is available according to the following steps. Concretely, the two hyperspectral images $X$ and $Y$ are firstly concatenated along the direction of image width to get the concatenation image $Z \in \mathbb{R}^{H \times (2 \cdot W) \times C}$. Next, the Hyperspectral Signal Identification by Minimum Error [39] is applied to compute the total number of all endmembers of two HSIs. Finally, we employ the Vertex Component Analysis [40] to calculate the endmember matrix $E$ from the concatenation image $Z$. We chose cosine similarity function (named as COS_SIM) to measure the difference between the predicted spectrum and the input center pixel spectrum. Cosine similarity computes the cosine of the angle between two vectors, and emphasizes the direction difference between two vectors not the absolute values. The closer the cosine similarity is to 1, the more parallel the two vectors are in direction, and vice versa. And the designed loss function can be described as follows:

$$L_{cos}(x, \hat{x}) = 1 - COS\_SIM(x, \hat{x}) = 1 - \frac{\sum_{i=1}^{C} x_i \cdot \hat{x}_i}{\sqrt{\sum_{i=1}^{C} x_i^2} \cdot \sqrt{\sum_{i=1}^{C} \hat{x}_i^2}} \quad (4)$$

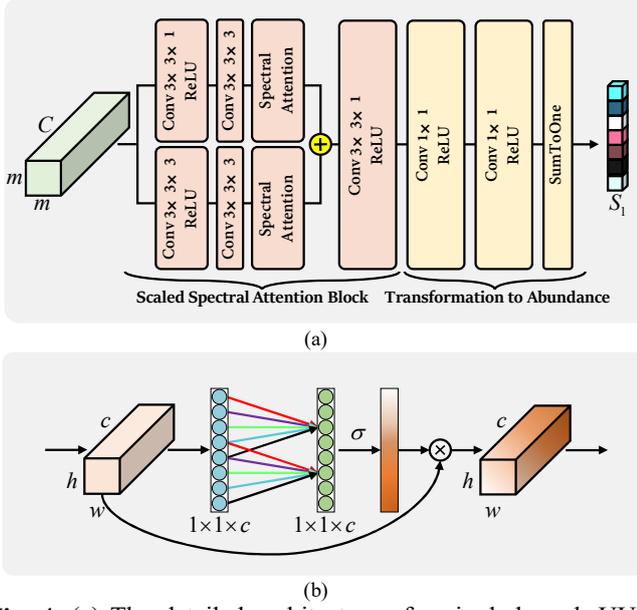

**Fig. 4.** (a) The detailed architecture of a single-branch UU-Module; (b) The Efficient Channel Attention block.

For the bi-temporal hyperspectral images, the united-unmixing module is actually a special partial-siamese network. The Scaled Spectral Attention Blocks of two hyperspectral images share the same weights for the purpose of balancing the similarity of bi-temporal inputs, and gain the same abundance for the unchanged pixels of the two HSIs. However, the weights in Transformation into Abundance Blocks are different and are updated independently to acquire more accurate united unmixing results in the case of two inputs with violent change.

Analogously, given the hyperspectral patch $Y_{patch} \in \mathbb{R}^{m \times m \times C}$ centered on the pixel $y \in \mathbb{R}^{C \times 1}$, the output abundance vector $S_2 \in \mathbb{R}^{K \times 1}$ of another single-branch UU-Module is depicted as:

$$S_2 = U_2(Y_{patch}) = F(C_7(C_6(C_3(Feature_Y))))$$
$$Feature_Y = ECA_1(C_{12}(C_{11}(Y_{patch}))) + ECA_2(C_{22}(C_{21}(Y_{patch}))) \quad (5)$$

where the $U_2$ refers to the second single branch of UU-Module with input as hyperspectral $Y$; $C_6$ and $C_7$ is the 2Dconv of Transformation into Abundance Block, differing from the $C_4$ and $C_5$ used in the first branch of UU-Module. The predictive spectrum of the center pixel $\hat{y} \in \mathbb{R}^{C \times 1}$ can be computed by:

$$\hat{y} = E \cdot S_2 \quad (6)$$

Consequently, the unmixing loss for the hyperspectral image $Y$ is expressed as follows:

$$L_{cos}(y, \hat{y}) = 1 - COS\_SIM(y, \hat{y}) = 1 - \frac{\sum_{i=1}^{C} y_i \cdot \hat{y}_i}{\sqrt{\sum_{i=1}^{C} y_i^2} \cdot \sqrt{\sum_{i=1}^{C} \hat{y}_i^2}} \quad (7)$$

*C. Temporal Correlation Module*

The TC-Module represents a new binary change analysis rule based on the bi-temporal abundance vectors, and then is backward to promote the multi-temporal unmixing process. The loss between the binary change detection result and the given binary pseudo-label describes the bias and error accumulation when binary change is computed from the unmixing result. It is noted that most of the unsupervised spectral unmixing networks are trained by minimizing the distance between the reconstructive spectrum and original spectrum to obtain the abundance[41, 42]. The shorter distance, however, does not mean better abundance when there is noise or spectral disturbance of imaging condition in the original HSI. And the proposed TC-Module places temporal correlation constraint on the UU-Module in the view of change detection, encouraging the multi-temporal abundances of the unchanged pixel to be consistent and that of changed pixel to be more accurate.

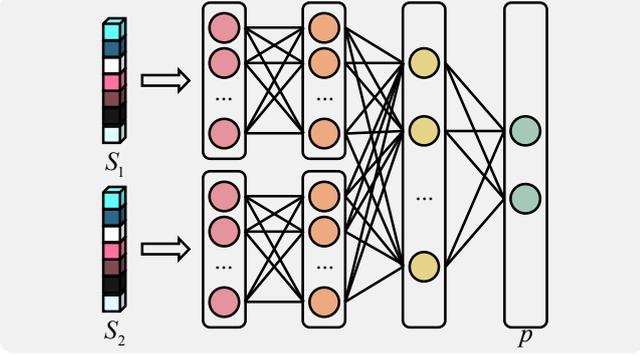

**Fig. 5.** The proposed TC-Module to learn the rules of detecting binary change information from two temporal abundance vectors automatically and adaptively.

As is shown in **Fig. 5**, the TC-Module comprises four fully connected layers. The pair of abundance vectors $S_1$ and $S_2$ are fed into the TC-Module to extract features from the abundance vectors separately, then are concatenated to extract the inter-temporal change features, finally are transformed into binary change information. The number of units in the output layer is set as two, where the first one refers to the binary classification probability of unchanged and the second one probability of changed. Considering the imbalanced samples between the changed and unchanged, focal loss [43] is a terrific choice to address the imbalanced problem and prevent the change detector or the binary classifier from being overwhelmed by the huge number of easy samples. Give a pair of abundance vectors $S_1$ and $S_2$, the output of the new change detection rule $Q(S_1, S_2)$ from TC-Module is expressed as follows:

$$\begin{aligned} p &= Q(S_1, S_2) = \text{SOFTMAX}(W_3 \cdot g(W_2 \cdot \text{Feature}_{XY} + B_2) + B_3) \\ \text{Feature}_{XY} &= \text{CONCAT}(g(W_{11} \cdot S_1 + B_{11}), g(W_{12} \cdot S_1 + B_{12})) \end{aligned} \quad (8)$$

where $W_{11}$ and $W_{12}$ are the weight matrices between the input layer and the first hidden layer of two input temporal abundance vectors; $B_{11}$ and $B_{12}$ represent the corresponding bias vectors; $W_2$ and $B_2$ are the weight matrix and bias vector between the first hidden layer and the second hidden layer, while $W_3$ and $B_3$ refer to that between the second hidden layer and the output layer. In addition, $g$ represents the ReLU activation function; CONCAT denotes the operation of concatenation, and SOFTMAX indicates the softmax operator that turns the extracted features into probability. The loss function we used is defined as follows:

$$L_{\text{fc}}(p, G) = -\alpha_t \cdot (1 - p_t)^\gamma \cdot \log(p_t) \quad (9)$$

$$p_t = \begin{cases} p & \text{if } G = 1 \\ 1 - p & \text{if } G = 0. \end{cases} \quad (10)$$

$$\alpha_t = \begin{cases} \alpha & \text{if } G = 1 \\ 1 - \alpha & \text{if } G = 0. \end{cases} \quad (11)$$

where $G$ refers to the given pseudo binary class label. And there is a weighting factor $\alpha \in [0,1]$ which is used to address the imbalanced samples. The modulating factor $(1 - p_t)^\gamma$ is able to adjust the proportion of easy samples' loss. For those easy samples, $p_t$ tends to be close to one. In such case, the bigger $\gamma$ is, the smaller the modulating factor $(1 - p_t)^\gamma$ is, thus decreasing the rate of the loss of the easy samples more greatly.

*D. The Scheme of Proposed BCG-Net*

The proposed BCG-Net integrated spectral unmixing and change detection in an unsupervised way for binary and multiclass change detection. For the purpose of quick convergence and optimization of the loss function, the two modules adopt warm up strategy firstly, and then are trained alternately. We adopted simple and widely used Change Vector Analysis (CVA) [44] and Expectation Maximization (EM) [45] to obtain pseudo binary labels. We can get change probability of all pixels from the output after training. A threshold value of 0.5 is employed to obtain the binary change map. For those changed pixels, each pixel is tagged as the class owning the maximum abundance value of the abundance vector. We then acquire the multiclass change map from the bi-temporal class comparison.

The Scheme of BCG-Net is depicted as **Algorithm 1**.

---

**Algorithm 1** Process of Training and Generating Binary and Multiclass Change Detection Result for BCG-Net

**Input:**
Hyperspectral image $X$, hyperspectral image $Y$, patch size $m$;
**Output:**
The binary and multiclass change detection map $I_b$ and $I_m$;
1: Calculate the multi-temporal endmember $E$;
2: Generate the training samples $\{x^{(j)}, j = 1, 2, \ldots, N\}$, $\{y^{(j)}, j = 1, 2, \ldots, N\}$, and pseudo binary labels $\{G^{(j)}, j = 1, 2, \ldots, N\}$ from pre-detection binary change map;
3: Warm up the UU-Module and TC-Module;
4: **while** $epoch < max\_epochs$ **do**
5:    Update the TC-Module by descending its stochastic gradient:

$$\nabla_{\theta_T} \sum_{j=1}^{N} \left[ \frac{1}{N} \cdot L_{\text{fc}}\left(Q\left(S_1^{(j)}, S_2^{(j)}\right), G^{(j)}\right) \right]$$

6:    Update the UU-Module by descending its stochastic gradient:

$$\nabla_{\theta_U} \sum_{j=1}^{N} \left[ L_{\cos}\left(x^{(j)}, E \cdot U_1\left(X_{\text{patch}}^{(j)}\right)\right) + L_{\cos}\left(y^{(j)}, E \cdot U_2\left(Y_{\text{patch}}^{(j)}\right)\right) \right.$$
$$\left. + \frac{1}{N} \cdot \omega \cdot L_{\text{fc}}\left(Q\left(U_1\left(X_{\text{patch}}^{(j)}\right), U_2\left(Y_{\text{patch}}^{(j)}\right)\right), G^{(j)}\right) \right]$$

where $\omega$ is the weight factor to leverage the losses of two modules.
7:  $epoch$++;
8: **end while**
9: Calculate the predicted binary change probability and get the binary change map $I_b$ with a threshold as 0.5;
10: Calculate the abundance maps of the two HSIs and acquire the multiclass change map $I_m$;
11: **return** $I_b$ and $I_m$;

---

## III. EXPERIMENTAL RESULTS AND ANALYSIS

*A. Datasets description*

The first dataset is the simulative Urban dataset designed for evaluating the proposed temporal correlation constraint on the performance of united unmixing and change detection. As the first row of **Fig. 6** shows, (a) is the original Urban dataset; (b) is the variant of (a) with simulative temporal changes; (c) and

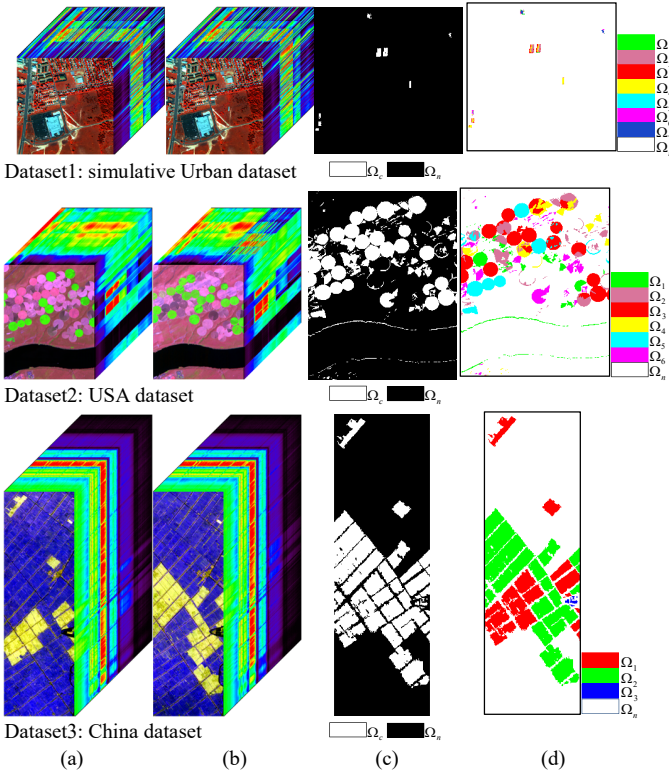

**Fig. 6.** From top to down are the simulative Urban dataset, USA dataset and China dataset. From left to right are (a) HSI $X$; (b) HSI $Y$; (c) Binary reference change map, $\Omega_n$ refers to the unchanged, $\Omega_c$ refers to the changed; (d) Multiclass reference change map, $\Omega_n$ is the unchanged and $\{\Omega_i\}_{i=1,2,\ldots 7}$ represent the different change classes.

TABLE I
DETAILS OF MULTICLASS CHANGE INFORMATION FOR THREE DATASETS

| NO. of change class | Number of change pixels | | |
|---|---|---|---|
| | Urban | USA | China |
| 1 | 29 | 1850 | 6559 |
| 2 | 169 | 2905 | 11613 |
| 3 | 119 | 5896 | 105 |
| 4 | 182 | 1524 | |
| 5 | 4 | 2744 | |
| 6 | 38 | 1808 | |
| 7 | 49 | | |

TABLE II
HYPERPARAMETERS FOR THREE DATASETS

| Hyperparameter | Urban | USA | China |
|---|---|---|---|
| $\alpha$, $\gamma$ | 2, 0.25 | 2, 0.5 | 1, 0.25 |
| $\omega$ | 1 | 1 | 20 |
| Learning rate | | 0.001 | |
| Weight decay rate | | 0.001 | |
| Batch size | | 64 | |
| Patch size $m$ | | 7 | |
| Epoch (whole training) | | 200 | |

(d) refer to binary and multiclass reference change map, respectively. The original Urban data was collected on Copperas Cove, USA, in 1995. The image size is 307 × 307, with 162 spectral bands. There are a great deal of trees, grasses and artificial architectures in the scene and four endmember materials were used in this experiment, including Asphalt, Grass, Tree, and Roof, respectively. Based on the original Urban data, several blocks were selected and inserted into another area to create changes of disappearance, insertion and movement. Additionally, we also added white Gaussian noise with different levels of SNR values (i.e., SNR = 20, 30, 40, 50 dB) to model the spectral variance of imaging conditions. The details of multiclasses change information are depicted as the first column of TABLE I with seven changes.

The second hyperspectral dataset is a real-world dataset, named as USA dataset. As is shown in the second row of **Fig. 6**, (a) is acquired on May 1, 2004, and (b) on May 8, 2007 by Hyperion in Hermiston city, USA in an irrigated agricultural field; (c) and (d) are the binary and multiclass reference change map, separately. The land cover types involve irrigated fields, soil, grassland, river, cultivated land, and building. The image covers a size of 307 × 241, with 154 spectral bands. TABLE I (second column) displays the detailed change information.

The third one, China dataset, is the other real dataset widely used for HMCD. **Fig. 6** (a) (b) of the third row present the bi-temporal hyperspectral datasets shot on May 3, 2006 and April 23, 2007 by Hyperion, in the city of Yuncheng, Jiangsu, China over farmland area. (c) and (d) are the binary and multiclass reference change maps, individually. The image comprises a size of 450 × 140, and 155 spectral bands. There are only three change classes in this dataset, as TABLE I (third column) shows.

## B. Experimental setting

We implemented our method by Pytorch and conducted experiments on a single NVIDIA RTX 2080 TI GPU. And He-normal [46] way was selected as initialization of the network, and Adam [47] as optimizer with L2 regularization, with the learning rate and weight decay rate both fixed as 0.001. The batch size is set as 64 and the patch size is equal to 7. The experiments are randomly repeated 3 times with random training data. The detailed hyperparameters settings are presented in TABLE II. For the pseudo binary labels, concretely, 2048 samples (2.17% of total pixel number, 1648 unchanged, 400 changed) from pre-detection result of EM are chose for the simulative Urban dataset with different levels of SNR values, respectively, and 9216 samples (12.26% of whole pixel number, 6144 unchanged, 3072 changed) from pre-detection result of EM for the USA dataset, and 12288 samples (19.50% of total pixel number, 8192 unchanged, 4096 changed) from pre-detection result of CVA for the China dataset.

Overall accuracy (OA) and the Kappa coefficient were calculated for quantitative evaluation of both binary change detection and multiclass change detection. To validate the effectiveness of proposed method, eight classical and state-of-the-art methods were conducted for comparison, with four binary change detection methods, including CVA [44], Iterative Slow Feature Analysis (ISFA) [48, 49], GETNET (without unmixing), as well as GETNET (with unmixing) [50], and another four multiclass change detection methods, Compressed Change Vector Analysis ($C^2VA$) [51], MSU [30], PUC [26], as well as (Re3FCN) [31] included. The multi-temporal principle component analysis is first applied to the hyperspectral images, and the main components are used for binary change detection for ISFA. GETNET proposes an end-to-end 2-D convolutional neural network framework for binary change detection, constructing a novel mixed affinity matrix for cross channel features learning. GETNET (with unmixing) integrates the pixel spectral information and sub-pixel abundance information to build the mixed affinity matrix and extract multi-source information. Noted that the number of changed class is provided for the Re3FCN as well as C$^2$VA as priori knowledge. Among all comparative methods, GETNET (with unmixing), MSU, PUC, Re3FCN all introduce unmixing or endmembers for change detection, which are opted specially for comparison. And GETNET (with unmixing), PUC and proposed method share the same endmember matrix extracted from the hyperspectral data for the sake of justice.

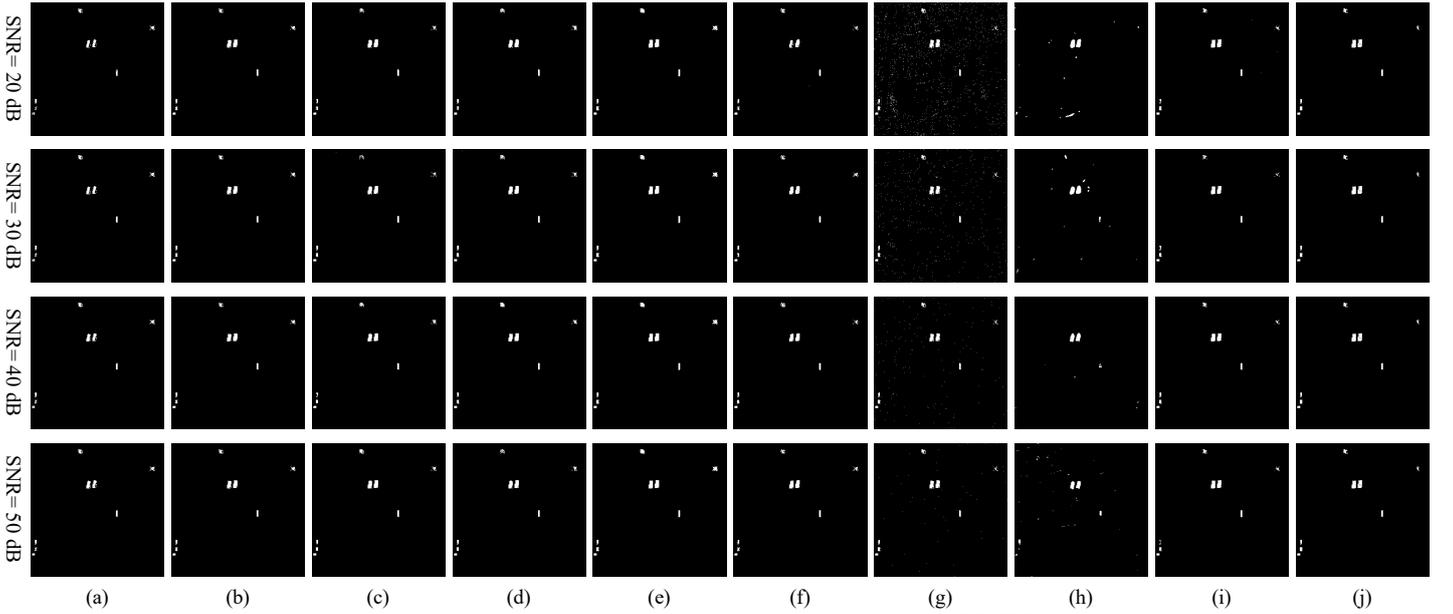

Fig. 7. The binary change detection of simulated Urban dataset. (a) CVA, (b) ISFA, (c) GETNET (without unmixing), (d) GETNET (with unmixing), (e) C$^2$VA, (f) MSU, (g) PUC, (h) Re3FCN, (i) BCG-Net, (j) Reference of binary change detection.

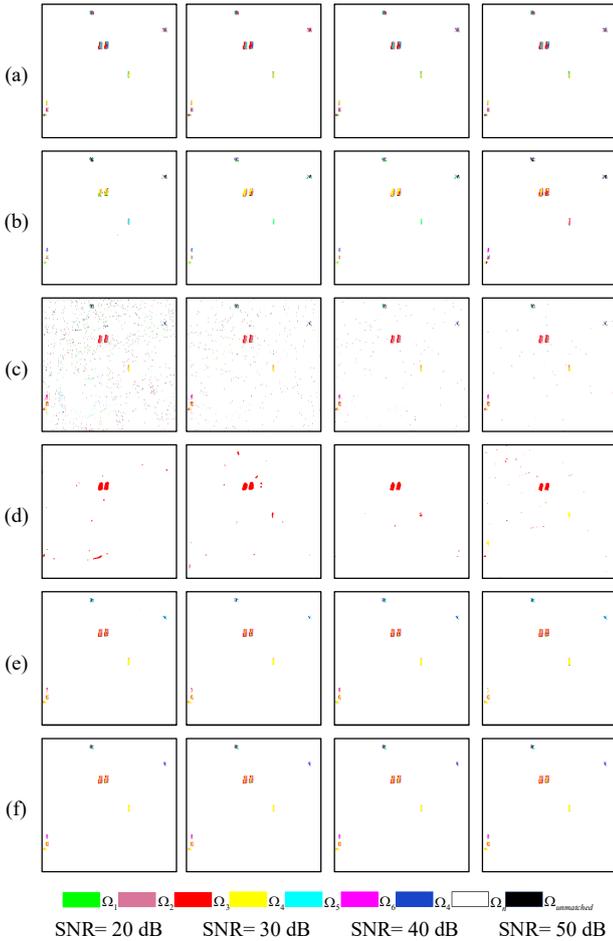

Fig. 8. The multiclass change detection of simulative Urban dataset. (a) C$^2$VA, (b) MSU, (c) PUC), (d)Re3FCN, (e)BCG-Net, (f) Reference of multiclass change detection.

### C. Change detection Results on Simulative Urban Datasets

**Fig. 7** depicts the binary change detection results on the simulative Urban datasets. As SNR decreases, the amount of noise in hyperspectral image grows. From the binary change detection results with SNR as 20db, all methods can detect most of changes except CVA, MSU and Re3FCN. And there is a great deal of noise in the result of PUC, which reduces greatly when the SNR of dataset increases. And more changes are detected in the results of CVA and MSU and less false alarm is found in the result of Re3FCN with higher SNR value. Among all the result of different comparative methods under variant SNR values, our proposed BCG-Net acquires the best binary change maps and is immune to the noise.

The multiclass change detection maps of proposed BCG-Net with regard to C$^2$VA, MSU, PUC, Re3FCN are shown in **Fig. 8**. Different kinds of change classes are displayed by different colors, areas of no change are in white, and those change classes that cannot match with the reference are in black. Noted that there are seven change classes, which brings huge challenge for multiclass change detection. As shown in **Fig. 8** (a), It is obvious that the result of $C^2$VA under the four SNR levels are not consistent with the reference. The results of PUC (**Fig. 8** (c)) are severely impacted by the noise. And Re3FCN does not work very well on this dataset due to small number of samples and numerous changes classes. Compared with these comparative methods, the multiclass change map of proposed BCG-Net is the most similar with the reference, performing well even when SNR is as low as 20db, and not affected by the small number of pixels of different changes class. Notably, UU-Module extracts both spectral and spatial features from the input and the TC-Module places temporal correlation constraint on the spectral unmixing from the point of change detection, which contribute to the noise-free change detection result.

TABLE III
QUANTITATIVE ASSESSMENT ON THE SIMULATIVE URBAN DATASET UNDER DIFFERENT SNR VALUES

| SNR | 20db | | | | 30db | | | |
|---|---|---|---|---|---|---|---|---|
| Method | Two-class | | Multiclass | | Two-class | | Multiclass | |
| | OA | Kappa | OA | Kappa | OA | Kappa | OA | Kappa |
| CVA | 0.9976 | 0.7855 | | | 0.9978 | 0.8100 | | |
| ISFA | 0.9990 | 0.9159 | | | 0.9990 | 0.9160 | | |
| GETNET (without unmixing) | 0.9986 | 0.8890 | | | 0.9985 | 0.8811 | | |
| GETNET (with unmixing) | 0.9985 | 0.8789 | | | 0.9989 | 0.9133 | | |
| $C^2$VA | 0.9989 | 0.9205 | 0.9947 | 0.6012 | 0.9990 | 0.9221 | 0.9947 | 0.6075 |
| MSU | 0.9985 | 0.8698 | 0.9964 | 0.6876 | 0.9989 | 0.9087 | 0.9959 | 0.6678 |
| PUC | 0.9852 | 0.4307 | 0.9834 | 0.3626 | 0.9942 | 0.6607 | 0.9924 | 0.5545 |
| Re3FCN | 0.9951 | 0.5777 | 0.9932 | 0.4240 | 0.9952 | 0.5851 | 0.9933 | 0.4212 |
| BCG-Net (ours) | **0.9992** | **0.9324** | **0.9986** | **0.8901** | **0.9991** | **0.9263** | **0.9987** | **0.8909** |
| SNR | 40db | | | | 50db | | | |
| Method | Two-class | | Multiclass | | Two-class | | Multiclass | |
| | OA | Kappa | OA | Kappa | OA | Kappa | OA | Kappa |
| CVA | 0.9982 | 0.8491 | | | 0.9982 | 0.8512 | | |
| ISFA | 0.9990 | 0.9163 | | | 0.9990 | 0.9163 | | |
| GETNET (without unmixing) | 0.9988 | 0.9017 | | | 0.9990 | 0.9205 | | |
| GETNET (with unmixing) | 0.9989 | 0.9152 | | | 0.9988 | 0.9017 | | |
| $C^2$VA | 0.9990 | 0.9222 | 0.9946 | 0.5994 | 0.9989 | 0.9215 | 0.9946 | 0.5989 |
| MSU | 0.9986 | 0.8856 | 0.9964 | 0.6934 | 0.9989 | 0.9109 | 0.9963 | 0.7008 |
| PUC | 0.9975 | 0.8183 | 0.9957 | 0.6868 | 0.9986 | 0.8893 | 0.9968 | 0.7459 |
| Re3FCN | 0.9961 | 0.6204 | 0.9943 | 0.4374 | 0.9960 | 0.6524 | 0.9937 | 0.4463 |
| BCG-Net (ours) | **0.9993** | **0.9463** | **0.9989** | **0.9098** | **0.9991** | **0.9316** | **0.9987** | **0.8973** |

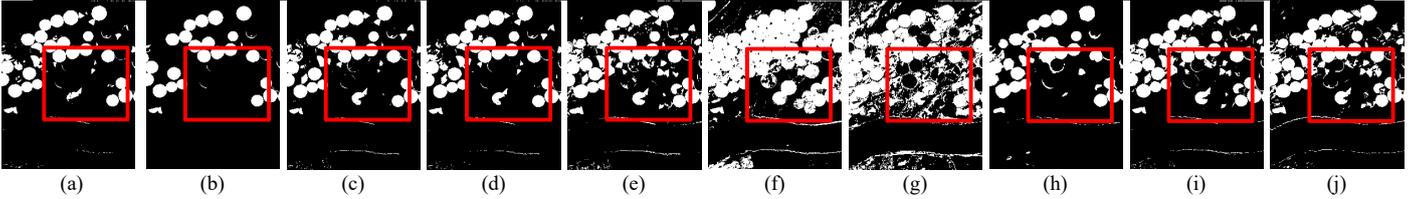

Fig. 9. The binary change detection of USA dataset. (a) CVA, (b) ISFA, (c) GETNET (without unmixing), (d) GETNET (with unmixing), (e) $C^2$VA, (f) MSU, (g) PUC, (h) Re3FCN, (i) BCG-Net, (j) Reference of binary change detection.

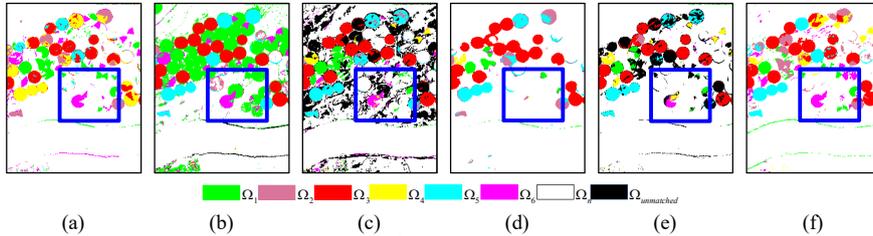

Fig. 10 The multiple change detection of USA dataset. (a) $C^2$VA, (b) MSU, (c) PUC), (d)Re3FCN, (e)BCG-Net, (f) Reference.

TABLE III summarizes the quantitative assessment of binary and multiclass change detection results. The maximum is marked in bold, and the second-best value is underlined. It is found that the proposed BCG-Net acquires the best OA and Kappa coefficient of both binary and multiclass change detection result under all SNR values. For binary change detection, the ISFA and $C^2$VA get the second-best OA and Kappa coefficient with different SNR values, respectively. As for multiclass change detection, the MSU obtains second-best OA and Kappa coefficient when SNR equals to 20db, 30db and 40db, separately. The PUC, however, obtains second-best OA and Kappa coefficient with SNR equal to 50db. It is worthy to mention that the second-best Kappa coefficient is much lower than the top one acquired by BCG-Net for multiclass change detection, confirming the superiority of proposed method.

*D. Change detection Results on USA dataset*

The binary change detection result of proposed BCG-Net on USA dataset as well as eight comparison methods are presented in **Fig. 9**. Compared with the binary reference change map

TABLE IV
QUANTITATIVE ASSESSMENT ON THE USA DATASET

| Method | Two-class | | Multiclass | |
|---|---|---|---|---|
| | OA | Kappa | OA | Kappa |
| CVA | 0.9272 | 0.7670 | | |
| ISFA | 0.9023 | 0.6716 | | |
| GETNET(without unmixing) | 0.9289 | 0.7764 | | |
| GETNET (with unmixing) | 0.9332 | 0.7897 | | |
| C²VA | **0.9564** | **0.8749** | **0.8666** | **0.6586** |
| MSU | 0.8132 | 0.5752 | 0.7140 | 0.4424 |
| PUC | 0.7825 | 0.4975 | 0.6945 | 0.4008 |
| Re3FCN | 0.9131 | 0.7255 | 0.8592 | 0.5921 |
| BCG-Net (ours) | 0.9546 | 0.8662 | 0.8456 | 0.5940 |

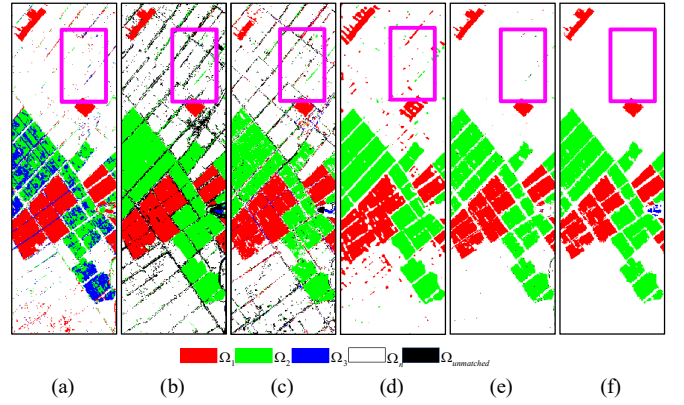

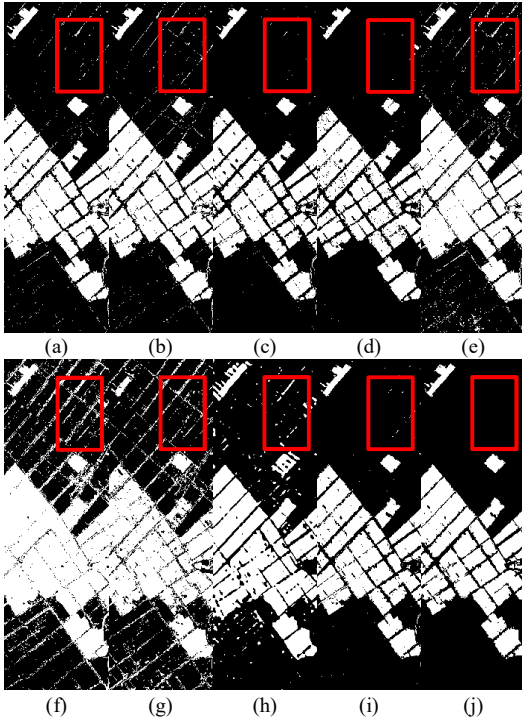

**Fig. 11.** The binary change detection of China dataset. (a) CVA, (b) ISFA, (c) GETNET (without unmixing), (d) GETNET (with unmixing), (e) C²VA, (f) MSU, (g) PUC, (h) Re3FCN, (i) BCG-Net, (j) Reference .

(**Fig. 9** (j)), it is obvious that there are larger white areas in the binary maps of MSU and PUC, revealing plenty of false positive changes. And there are some omitted changed areas in result of CVA, ISFA and Re3FCN. By contrast, GETNET, C²VA and proposed BCG-Net could detect most of changes accurately.

**Fig. 10** shows the multiclass change detection maps. Lots of unmatched black areas can be found in the result of PUC (**Fig. 10** (c)), where the before and after abundance comparison brings about far more change indications than the reference. By contrast, no black area is found in the multiclass change maps of C²VA and Re3FCN due to the provided prior information of the number of change classes, avoiding the redundant unmatched change classes. As observed, most of the green area in the result of MSU is unchanged in effect. It is found that BCG-Net can detect most change classes correctly, despite of a piece of unmatched area. The constraint of temporal correlation supported by pseudo binary labels is able to optimizes the united unmixing to encourage the coherence of the abundances of those unchanged objects and improve the accuracy of the abundances of those change objects.

TABLE IV reports the quantitative assessment of the results on USA dataset. It is observed that C²VA achieves the largest OA and Kappa coefficient as 0.9564 and 0.8749 for binary and as 0.8666 and 0.6586 for multiclass change detection results, respectively; BCG-Net gains second-best quantitative assessment performance as OA 0.9546 and Kappa 0.8662 for binary change detection and 0.5940 as Kappa for multiclass change detection, and the Re3FCN acquires the second-best OA as 0.8592 for the multiclass change detection result. The performance of PUC is not optimistic, far behind the best OA and Kappa coefficient. Although BCG-Net gains the second-best performance, BCG-Net is an unsupervised method for both binary and multiclass change detection, acquiring a relatively nice effect for both binary and multiclass change maps.

*E. Change detection Results on China dataset*

**Fig. 11** and **Fig. 12** are the binary and multiclass change detection results on China dataset. As shown in **Fig. 11**, GETNET and proposed BCG-Net acquire nice performance on the binary change results, with most changes detected and little noise. It is observed that the binary change result of BCG-Net is not affected by the noise in the result of pre-detection algorithm CVA. Many stripes misclassified as changes can be found in the change maps of C²VA, MSU and PUC, weakening the performance of multiclass change result further.

**Fig. 12.** The multiclass change detection of China dataset. (a) C²VA, (b) MSU, (c) PUC), (d)Re3FCN, (e)BCG-Net, (f) Ground truth map of multiclass change detection.

TABLE V
QUANTITATIVE ASSESSMENT ON CHINA DATASET

| Method | Two-class | | Multiclass | |
|---|---|---|---|---|
| | OA | Kappa | OA | Kappa |
| CVA | 0.9548 | 0.8926 | | |
| ISFA | 0.9575 | 0.8996 | | |
| GETNET(without unmixing) | 0.9584 | 0.8974 | | |
| GETNET (with unmixing) | 0.9626 | 0.9076 | | |
| C²VA | 0.9364 | 0.8534 | 0.8627 | 0.7238 |
| MSU | 0.8432 | 0.6713 | 0.8407 | 0.7121 |
| PUC | 0.8650 | 0.6997 | 0.8612 | 0.7280 |
| Re3FCN | 0.9306 | 0.8381 | 0.9305 | 0.8536 |
| BCG-Net (ours) | **0.9664** | **0.9184** | **0.9660** | **0.9249** |

TABLE VI
THE MSE COMPARISON OF ABUNDANCE MAP FOR SPECTRAL UNMIXING WITH AND WITHOUT PROPOSED
TEMPORAL CORRELATION CONSTRAINT

|   | SNR | 20db | | | 30db | | | 40db | | | 50db | | |
|---|---|---|---|---|---|---|---|---|---|---|---|---|---|
|   | Method | FCLS | UU-Module | BCG-Net | FCLS | UU-Module | BCG-Net | FCLS | UU-Module | BCG-Net | FCLS | UU-Module | BCG-Net |
| $X$ | Asphalt | 0.0595 | 0.0067 | **0.0064** | 0.0595 | 0.0071 | **0.0064** | 0.0595 | 0.0071 | **0.0059** | 0.0595 | 0.0077 | **0.0058** |
|   | Grass | 0.0266 | **0.0043** | **0.0043** | 0.0266 | 0.0050 | **0.0044** | 0.0266 | 0.0049 | **0.0040** | 0.0266 | 0.0051 | **0.0041** |
|   | Tree | 0.0219 | **0.0019** | **0.0019** | 0.0219 | 0.0031 | **0.0019** | 0.0219 | 0.0026 | **0.0018** | 0.0219 | **0.0018** | 0.0020 |
|   | Roof | 0.0793 | 0.0022 | **0.0020** | 0.0793 | 0.0028 | **0.0021** | 0.0793 | 0.0027 | **0.0021** | 0.0793 | **0.0021** | **0.0021** |
|   | Average | 0.0468 | 0.0038 | **0.0036** | 0.0468 | 0.0045 | **0.0037** | 0.0468 | 0.0043 | **0.0035** | 0.0468 | 0.0042 | **0.0035** |
| $Y$ | Asphalt | 0.0600 | 0.0074 | **0.0073** | 0.0598 | 0.0075 | **0.0070** | 0.0598 | 0.0068 | **0.0060** | 0.0598 | 0.0075 | **0.0062** |
|   | Grass | 0.0269 | 0.0054 | **0.0052** | 0.0266 | 0.0054 | **0.0048** | 0.0266 | 0.0050 | **0.0041** | 0.0266 | 0.0048 | **0.0040** |
|   | Tree | 0.0222 | **0.0023** | 0.0024 | 0.0220 | 0.0031 | **0.0021** | 0.0220 | 0.0022 | **0.0019** | 0.0220 | **0.0018** | 0.0019 |
|   | Roof | 0.0801 | **0.0025** | 0.0026 | 0.0800 | 0.0029 | **0.0022** | 0.0800 | 0.0026 | **0.0021** | 0.0800 | 0.0022 | **0.0021** |
|   | Average | 0.0473 | **0.0044** | **0.0044** | 0.0471 | 0.0047 | **0.0040** | 0.0471 | 0.0042 | **0.0035** | 0.0471 | 0.0040 | **0.0036** |

For multiclass change maps, many unmatched areas in black in **Fig. 12** (b) and (c) indicate redundant change classes detected by MSU and PUC. The change class in blue shown in **Fig. 12** (a) of C2VA is not matched with the reference. By contrast, there are little noise and black unmatched area in the multiclass change map of BCG-Net.

For quantitative evaluation, TABLE V lists the OA and Kappa coefficient on binary and multiclass change results. Concretely, for the assessment of binary change detection, the OA and Kappa coefficient of BCG-Net rank first, equal to 0.9664 and 0.9184, respectively. GETNET (with unmixing) wins second-best as OA 0.9626 and Kappa coefficient 0.9076, separately. And the result of MSU gets the worst performance, resulting from voluminous false alarm. For the assessment of multiclass change detection, the top of OA and Kappa coefficient are 0.9660 and 0.9249 respectively, attained by BCG-Net. Re3FCN wins second-best OA and Kappa coefficient as 0.9305 and 0.8536, separately. To conclude, BCG-Net gains best OA and Kappa coefficient of both binary and multiclass change results, exhibiting the effectiveness of proposed method.

## IV. Discussions

On this section, the effect of temporal correlation constraint on the spectral unmixing is firstly discussed. Then the discussion on the effect of temporal correlation module on the binary and multiclass change detection is provided.

### A. The Effect of Temporal Correlation Constraint on Spectral Unmixing

Here, the unmixing performance comparison on the simulative Urban dataset is represented to test the effect of the proposed temporal correlation constraint. For sake of fairness, we use the result of UU-Module as comparison without temporal correlation. And we also test the abundance map obtained by fully constrained lest square (FCLS) [52]. The metric to evaluate the abundance maps is depicted as:

$$\text{MSE} = \frac{1}{K}\sum_{i=1}^{K} \left\| S_j^{(i)} - \hat{S}_j^{(i)} \right\|^2, j = 1, 2 \quad (12)$$

where $K$ is number of multi-temporal endmembers; $S_1^{(i)}$ and $S_2^{(i)}$ refer to abundance map of the endmember $i$ of HSI $X$ and HSI $Y$; $\hat{S}_1^{(i)}$ and $\hat{S}_2^{(i)}$ are the corresponding reference abundance maps.

The MSE comparison result is represented in TABLE VI. Best results are in bold. As can be seen, the MSE of UU-Module and BCG-Net are far lower than that of FCLS for both HSI $X$ and $Y$ under four different SNR values. It might be contributed to the powerful feature extraction ability of designed united unmixing network, where spectral and spatial information are taken into consideration at the same time. Moreover, BCG-Net obtains lower abundance MSE than UU-Module does for most of the endmembers of two HSIs at different SNR values, which confirms that the temporal correlation constraint really works to improve the performance of spectral unmixing. For BCG-Net, the designed TC-Module is able to constraint the unmixing result from the view of binary change detection. Generally, for the unchanged pixels, the abundances of them at two temporal HSIs are probably similar with each other. Likewise, the abundances of the changed pixels at two temporal HSIs are probable different with each other. With the stimulation of the temporal correlation constraint, the abundances of those unchanged pixels are encouraged to be consistent and that of changed pixels be more accurate, contributing to greater multi-temporal unmixing results. Besides, the abundance MSE of HSI $Y$ is higher than the one of HSI $X$ when SNR is equivalent to 20db, 30db. This is because the massive noise resulting from low SNR undermines the quality of hyperspectral images and further influences the performance of unmixing. And the abundance MSE of all unmixing result improves with higher SNR value.

**Fig. 13** shows the visual comparison of abundance maps on simulative Urban dataset under SNR as 20db. It is obvious that the abundance maps of UU-Module (**Fig. 13**, the second row) and BCG-Net (**Fig. 13**, the third row) are closely similar with the reference (**Fig. 13**, the fourth row). FCLS, however, performs not well especially in the abundance maps on the Asphalt (**Fig. 13** (a), (e), the first row) and the Roof (**Fig. 13** (d), (h), the first row), compared with the reference. For better visualization of the difference between the abundance results of UU-Module and BCG-Net, the residual error between the estimated abundance and the reference is computed according to:

$$\text{R}_{\text{MSE}} = \frac{1}{K}\sum_{i=1}^{K} (S_j^{(i)} - \hat{S}_j^{(i)})^2, j = 1, 2 \quad (13)$$

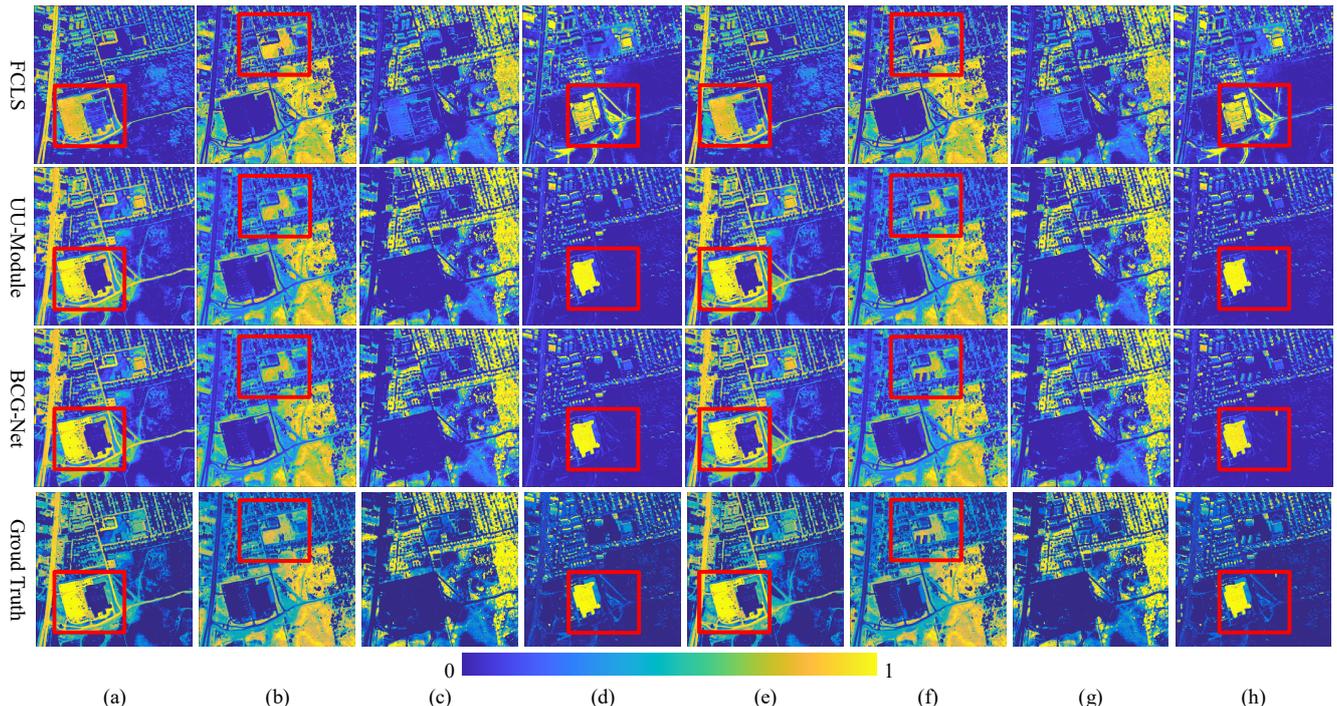

**Fig. 13.** The visual comparison of estimated abundance maps on simulative Urban dataset with SNR equals to 20db. From top to bottom are abundance maps of FCLS, UU-Module, BCG-Net and ground truth. From left to right are the estimated abundance maps of HSI $X$ on the (a) Asphalt, (b) Grass, (c) Tree, and (d) Roof, as well as the abundance maps of HSI $Y$ on the (e) Asphalt, (f) Grass, (g) Tree, and (h) Roof.

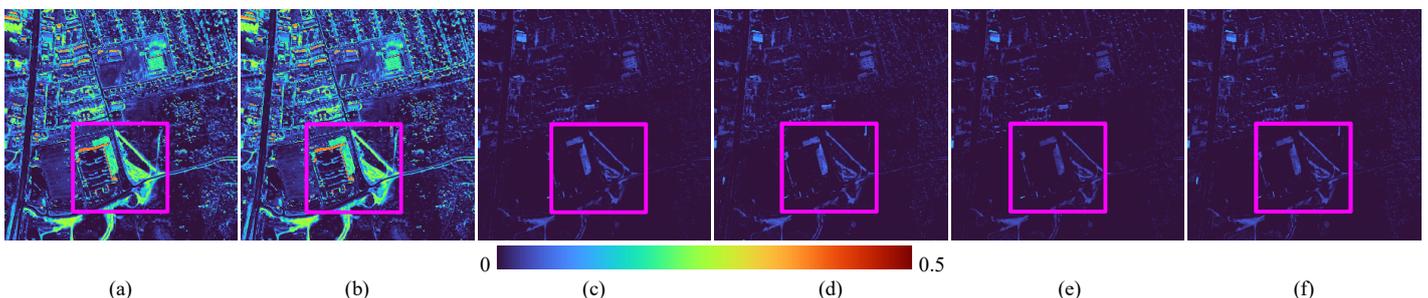

**Fig. 14.** The visual comparison of residual error maps of estimated abundance maps on simulative Urban dataset with SNR equals to 20db. From top to bottom are abundance maps of FCLS, UU-Module, BCG-Net and ground truth. From left to right are the residual error maps of FCLS on (a) HSI $X$, (b) HSI $Y$, the residual error maps of UU-Module on (c) HSI $X$, (d) HSI $Y$, and residual error maps of BCG-Net on (e) HSI $X$ and (f) HSI $Y$.

**Fig. 14** represents the residual error maps of estimated abundance maps from FCLS, UU-Module and BCG-Net on simulative Urban dataset with SNR equals to 20db. From the marked pinked frame, it is observed that the error of BCG-Net is lower than that of UU-Module, both of which are largely lower than the residual error acquired by FCLS. All in all, it is concluded that the proposed BCG-Net achieves more accurate abundance maps than UU-Module and FCLS does, demonstrating the effectiveness of proposed temporal correlation constraint on the spectral unmixing.

*B. The Effect of Temporal Correlation Module on Change Detection*

To test the effect of temporal correlation module on change detection, comparison experiment of proposed BCG-Net and the traditional post unmixing comparison is conducted on the three datasets. To be fair, the spectral unmixing method here both employ the proposed UU-Module, named as PUC(UU-Module). Additionally, the PUC with FCLS as spectral unmixing method is also taken for comparison, named as PUC(FCLS).

TABLE VII represents the quantitative comparison of binary and multiclass change results of these three methods. The greatest result is in bold. For all datasets, BCG-Net acquires much better OA and Kappa coefficient in both binary and multiclass change detection result than PUC(UU-Module) and PUC(FCLS) do. Compared with PUC(UU-Module), BCG-Net obtains more 16% around rise on the OA of both binary and multiclass change results for USA dataset, and 3% around growth on that of China dataset. Besides, for the Kappa coefficient of both binary and multiclass result, BCG-Net acquires a more 22% to 58% increasement than PUC(UU-Module) does especially for simulative Urban dataset and USA dataset, and 2% and 5% growth on the China dataset. Compared with PUC(FCLS), BCG-Net obtains a more 10% to 17% rise on the OA of both binary and multiclass results for USA and China datasets, and more 19% to 52% increasement on the Kappa coefficient of that for all the three datasets.

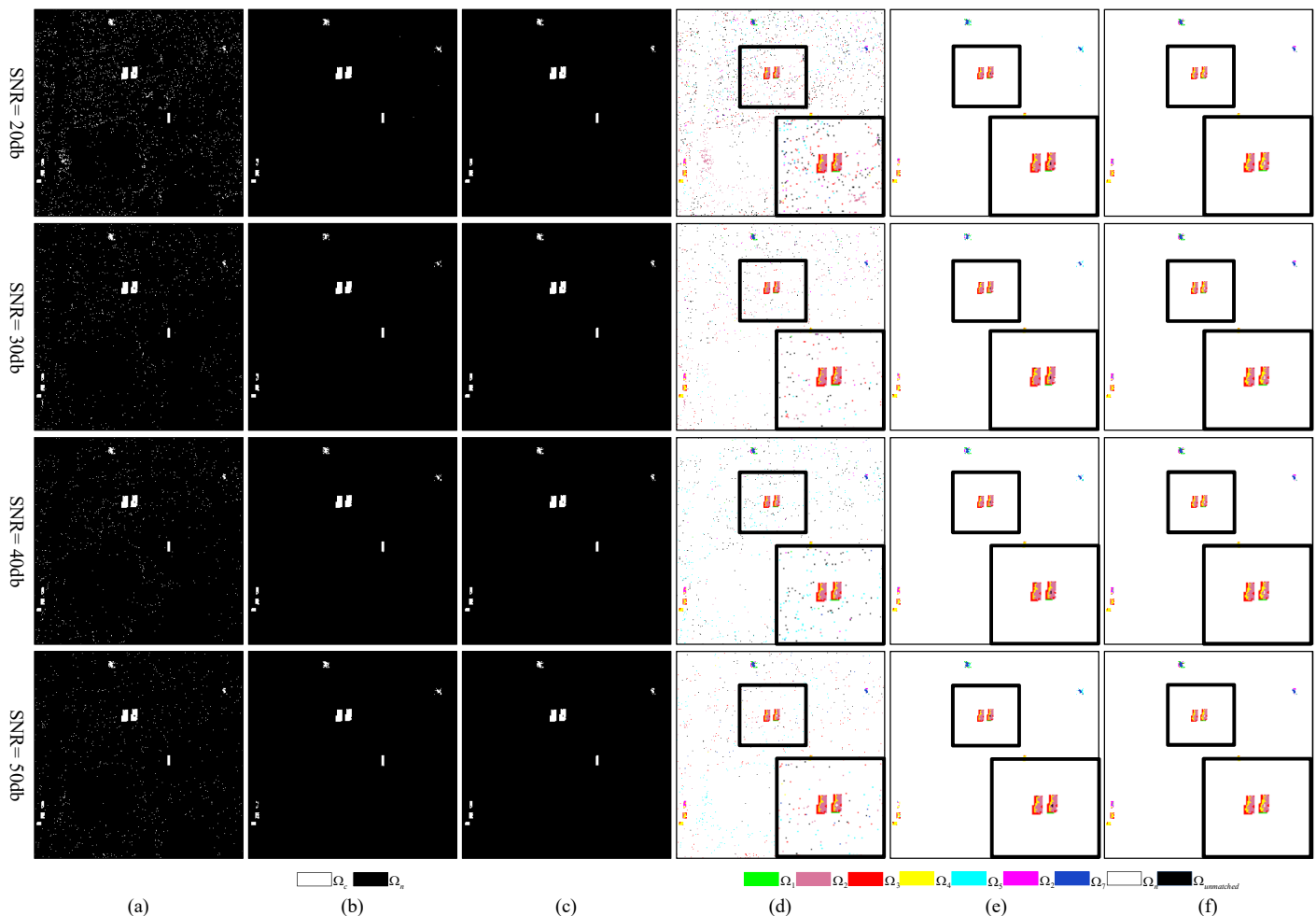

**Fig. 15.** The visual comparison of change detection result on the simulative Urban dataset between the traditional PUC(UU-Module) and the proposed BCG-Net. From left to right are binary change maps of (a) traditional PUC, (b) proposed BCG-Net, (c) binary ground truth, and multiclass change maps of (d) traditional PUC(UU-Module), (e) proposed BCG-Net, (f) multiclass ground truth, separately. From top to down are results on Simulative Urban dataset under SNR as 20db, 30db, 40db and 50db, respectively.

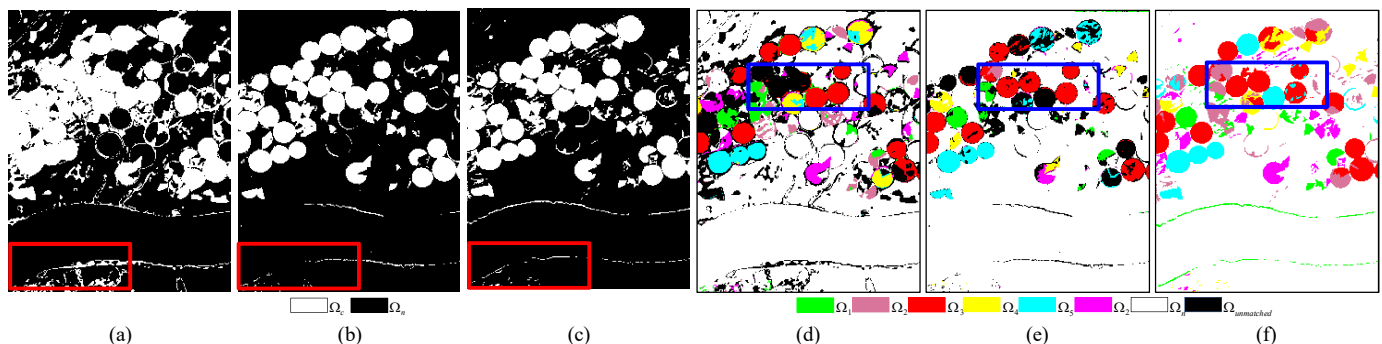

**Fig. 16.** The visual comparison of change detection result on the USA dataset between the traditional PUC(UU-Module) and the proposed BCG-Net. From left to right are binary change maps of (a) traditional PUC(UU-Module), (b) proposed BCG-Net, (c) binary ground truth, and multiclass change maps of (d) traditional PUC(UU-Module), (e) proposed BCG-Net, (f) multiclass ground truth.

It is noted that PUC(FCLS) obtains better performance than PUC(UU-Module) does for simulative Urban dataset and USA dataset, which indicates that the traditional PUC method does not always gain better change detection result with better unmixing result. The reason can be summarized as follows. Traditional PUC method usually adopts such a rule that each pixel is firstly assigned the class holding the maximum of the abundance vector and then the bi-temporal class labels are compared to acquire binary change information. That is to say, the class with the maximum abundance value is regarded as the main component of the pixel. If the main components remain unchanged, the pixel is considered to be unchanged. Otherwise, the pixel is thought to be changed. However, this brief binary change detection rule is susceptible to subtle differences of abundance values. For example, a mixed pixel is composed of only two endmembers with equal abundances and keeps unchanged during two phases. The abundance components of these two endmembers, however, are not equal resulting from numerical computation of unmixing process, leading to the pixel altered. And the phenomenon is more common in the datasets with complicated background, such as simulative Urban dataset and USA dataset.

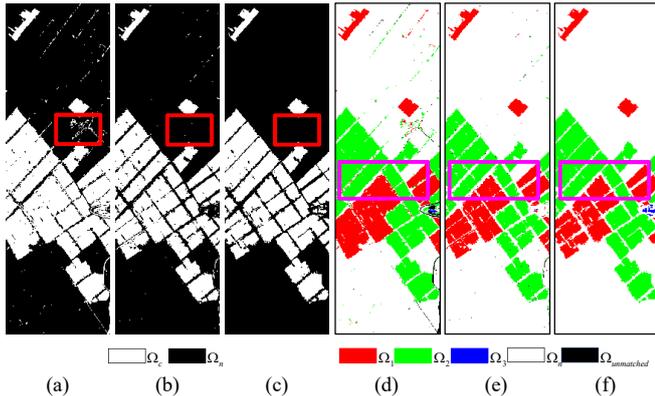

**Fig. 17.** The visual comparison of change detection result on the China dataset between the traditional PUC(UU-Module) and the proposed BCG-Net. From left to right are binary change maps of (a) traditional PUC(UU-Module), (b) proposed BCG-Net, (c) binary ground truth, and multiclass change maps of (d) traditional PUC(UU-Module), (e) proposed BCG-Net, (f) multiclass ground truth.

For the proposed BCG-Net, apart from casting a temporal correlation constraint on the UU-Module, TC-Module also plays a role of powerful binary change detector, seeking the exact relationship between the multi-temporal abundance vectors pair and the binary change information and bringing no false detection problem. Besides, the UU-Module and TC-Module are optimized alternately, eliminating the bias and accumulated errors from abundance result to change detection result. **Fig. 15** presents the visual comparison of binary and multiclass change maps of the traditional PUC (UU-Module) method and BCG-Net on simulative Urban dataset. Massive noise can be observed in the binary and multiclass change maps of PUC (UU-Module), as shown in **Fig. 15** (a), (d), whereas most of them disappear in the result of BCG-Net (**Fig. 15** (b), (e)). **Fig. 16** is another example of the effect of the TC-Module on the USA dataset. Compared with the result of PUC (UU-Module) in **Fig. 16** (a), BCG-Net (shown in **Fig. 16** (b)) can detect more accurate change area and less false alarms on the binary change maps. And for the multiclass change maps, as the blue frame marks, BCG-Net (shown in **Fig. 16** (e)) acquires more sound multi-class change detection results than PUC (UU-Module) (shown in **Fig. 16** (d)) does. This is because that the TC-Module encourages the UU-Module to acquire better unmixing result, which contributes to more accurate multiclass change detection result. From the comparison result of China dataset shown in **Fig. 17**, the noises in the red box of the binary change map of PUC (UU-Module) (**Fig. 17** (a)) are eliminated to some extent in the result of BCG-Net (**Fig. 17** (b)). And the multiclass change map of BCG-Net (**Fig. 17** (e)) is more similar with the reference than that of PUC (UU-Module) (**Fig. 17** (d)).

TABLE VII
QUANTITATIVE COMPARISON ON THE CHANGE DETECTION PERFORMANCE BETWEEN THE PUC(UU-MODULE) WITHOUT TEMPORAL CONSTRAINT AND BCG-NET WITH TEMPORAL CONSTRAINT

| Dataset | Model | Two-class | | Multiclass | |
|---|---|---|---|---|---|
| | | OA | Kappa | OA | Kappa |
| Urban (20db) | PUC(UU-Module) | 0.9779 | 0.3522 | 0.9774 | 0.3409 |
| | PUC(FCLS) | 0.9852 | 0.4307 | 0.9834 | 0.3626 |
| | BCG-Net | **0.9992** | **0.9324** | **0.9986** | **0.8901** |
| Urban (30db) | PUC(UU-Module) | 0.9902 | 0.5523 | 0.9896 | 0.5274 |
| | PUC(FCLS) | 0.9942 | 0.6607 | 0.9924 | 0.5545 |
| | BCG-Net | **0.9991** | **0.9263** | **0.9987** | **0.8909** |
| Urban (40db) | PUC(UU-Module) | 0.9911 | 0.5889 | 0.9906 | 0.5647 |
| | PUC(FCLS) | 0.9975 | 0.8183 | 0.9957 | 0.6868 |
| | BCG-Net | **0.9993** | **0.9463** | **0.9989** | **0.9098** |
| Urban (50db) | PUC(UU-Module) | 0.9916 | 0.5919 | 0.9911 | 0.5707 |
| | PUCFCLS) | 0.9986 | 0.8893 | 0.9968 | 0.7459 |
| | BCG-Net | **0.9991** | **0.9316** | **0.9987** | **0.8973** |
| USA | PUC(UU-Module) | 0.7926 | 0.5237 | 0.6769 | 0.3703 |
| | PUC (FCLS) | 0.7825 | 0.4975 | 0.6945 | 0.4008 |
| | BCG-Net | **0.9546** | **0.8662** | **0.8456** | **0.5940** |
| China | PUC(UU-Module) | 0.9393 | 0.8589 | 0.9383 | 0.8710 |
| | PUC(FCLS) | 0.8650 | 0.6997 | 0.8612 | 0.7280 |
| | BCG-Net | **0.9664** | **0.9184** | **0.9660** | **0.9249** |

In summary, with TC-Module, BCG-Net shows better performance on reducing the false alarm of binary change map and boosting the multiclass detection result.

## V. Conclusion

In this article, we propose an unsupervised hyperspectral multiclass change detection method named as BCG-Net to solve the problem of error accumulation and neglection of temporal correlation encountered by traditional methods. Instead of obtaining the binary and multiclass changes directly from the unmixing result like most previous methods, a novel temporal correlation constraint directed by pseudo binary labels is designed to boost the spectral unmixing process from the point of view of change detection, where the abundance of the unchanged pixels is encouraged to be more consistent and that of the changed pixels more accurate. Besides, we put forward a new rule based on neural network to build an effective relationship between the abundance pairs and the change information, bypassing the abundant false alarms traditional rule suffers from. The represented innovative strategy of iteratively optimizing the unmixing process and the change detection process provides a terrific solution to eliminate the accumulated error and bias from unmixing result to changed detection result. The qualitative and quantitative evaluation are conducted on three hyperspectral datasets to test the effectiveness of proposed method. In summary, the proposed binary change guided hyperspectral multiclass change detection network achieves competitive or even superior result on all tested datasets, demonstrating the validity of temporal correlation constraint on the binary and multiclass change detection results, as well as the multi-temporal spectral unmixing result.